\documentclass[final,english]{bullsrsl}[2022/06/15]
\usepackage[latin1]{inputenc}
\usepackage[T1]{fontenc}

\usepackage{natbib} 
\usepackage{graphicx}

\begin{document}
\title{Influence of mid-infrared Galactic bubble on surroundings: A case study on IRAS\,16489-4431}

\author[affil={1}, corresponding]{Ariful}{Hoque}
\author[affil={1}, corresponding]{Tapas}{Baug}
\author[affil={2}]{Lokesh}{Dewangan}
\author[affil={3}]{Ke}{Wang}
\author[affil={4}]{Tie}{Liu}
\author[affil={1}]{Soumen}{Mondal}
\affiliation[1]{S. N. Bose National Centre for Basic Sciences, Sector-III, Salt Lake, Kolkata-700 106, India}
\affiliation[2]{Astronomy \& Astrophysics Division, Physical Research Laboratory, Navrangpura, Ahmedabad 380009, India}
\affiliation[3]{Kavli Institute for Astronomy and Astrophysics, Peking University, Haidian District, Beijing 100871, People's Republic of China}
\affiliation[4]{Shanghai Astronomical Observatory, Chinese Academy of Sciences, 80 Nandan Road, Shanghai 200030, People's Republic of China}
\correspondance{ariful.hoque@bose.res.in, tapasbaug@bose.res.in}
\date{30th June 2023}
\maketitle


%

\begin{abstract}
We studied the influence of a massive star on a mid-infrared bubble and its surrounding gas in the IRAS\,16489-4431 star-forming region using multi-wavelength data.
The {\it Spitzer} mid-infrared band images revealed the shocked nature of the bubble. Analyses showed that the bubble is developed by a massive star owing
to its strong radiation pressure. Evidence of collected material along the edge of the bubble was noted by the cold gas tracer line observed using Atacama Millimeter/submillimeter Array (ALMA).
The presence of dense dust cores with bi-polar outflows and young stellar objects toward the collected material is suggestive of active star formation possibly influenced by the
expansion of the radiation driven bubble.
\end{abstract}


\section{Introduction}
Massive stars (M$_\ast$>8 M$_{\odot}$), even though fewer in number, play a pivotal role in determining the evolution and ecology of their host galaxies \citep{zinnecker2007}. They, in general, affect
their local environment throughout their life-time, by driving powerful jets and outflows, strong stellar winds, and by radiating substantial amounts of ultraviolet photons. Owing to such strong
stellar feedback, they could either influence the surrounding gas for the formation of the next generation of stars or disperse the natal environment into the interstellar
medium and halt further star formation \citep{deharveng2010}.

Surrounding medium of a massive star gets ionized because of strong ultraviolet radiation, and develops an ionized region (i.e., H{\sc ii} region). Supersonic expansion of H{\sc ii} regions develop
shock fronts \citep{stromgren1939} and induce compression in neighboring molecular gas that morphologically appears as a bubble-like structure. Such bubbles were widely identified
in {\it Spitzer} mid-infrared (MIR) 8$\mu$m band images \citep{churchwell2006, churchwell2007}. 

Expansion of the MIR bubble could lead to the formation of new stars at the periphery of the bubbles \citep{deharveng2010, dale2012}. The compressed gas at the periphery
of the bubbles may become gravitationally unstable and fragment into denser cores that can lead to the formation of stars \citep{elmegreen1977}.
Evidence of such star formation inferred by the presence of cluster of young stellar objects (YSOs), clumps and cores is found in several studies (see e.g., \citet{baug2016, baug2019, dewangan2020, das2017}).

\begin{figure}[t]
\centering
\includegraphics[width=\textwidth]{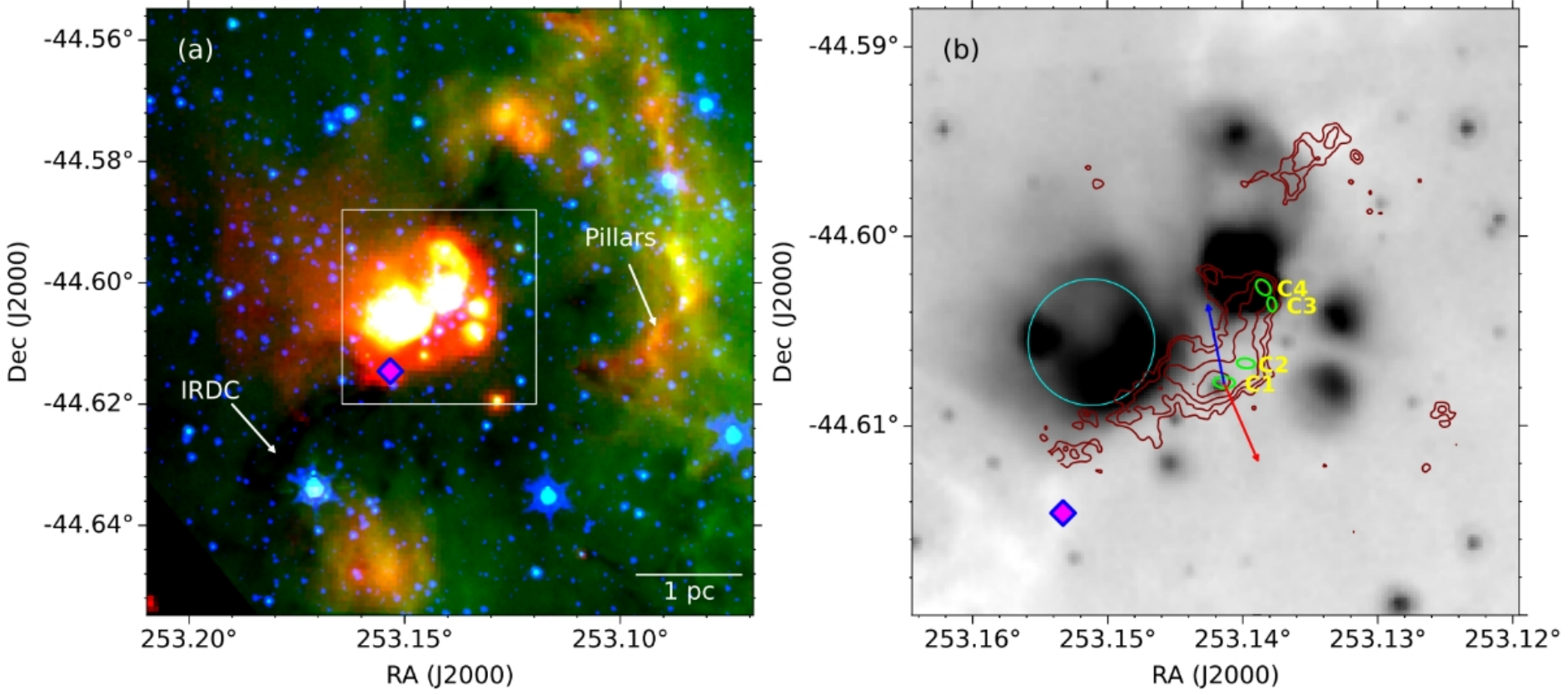}
\bigskip
\begin{minipage}{12cm}
\caption{(a) A three-color composite image (red: 70 $\mu$m, green: 8 $\mu$m, and blue: 4.5 $\mu$m) of the region. The white rectangle shows our target area for this study. A magenta diamond shows 
a Class {\sc ii} methanol maser source, and an IRDC and pillar-like structures are also labelled. (b) {\it Spitzer}-IRAC 8.0 $\mu$m image of target area overlaid with integrated H$^{13}$CO+ contours
in maroon drawn at [3, 5, 8, 10]$\times\sigma$ level. The cyan circle marks the MIR bubble. The cores are marked by ellipses and labelled (C1--C4). The red and blue arrows are
the SiO outflow lobes associated with C1 (Baug et al., in prep).}
\label{RGB}
\end{minipage}
\end{figure}

In this paper, we study one such region, IRAS\,16489-4431 where observations have shown the presence of a MIR bubble \citep{jayasinghe2019} and also cold mm-band cores
around its periphery. The bubble was identified using a citizen science survey on the three-color images made using {\it Spitzer Space Telescope} Infrared Array Camera (IRAC; spatial
resolution of $\sim$2$''$) and Multiband Imaging Photometer for Spitzer (MIPS; spatial resolution $\sim$6$''$) 24 $\mu$m band images.
The region is located at a near-kinematic distance of 3.26 kpc \citep{urquhart2018}. A three color composite image of the region is shown in Fig.\,\ref{RGB}(a). 
Presence of an infrared dark cloud (IRDC) and pillars can be observed in the region. Presence of dust cores at the periphery of the bubble makes this region 
important for exploring the influence of a massive star on its surrounding gas.
The paper is presented in the following manner. The data used in this paper are presented in Section\,\ref{sec:data}. The results of this study are described in Section\,\ref{sec:results}.
A discussion of the results is presented in Section\,\ref{sec:discussion}. Finally, we summarize the results in Section\,\ref{sec:summary}.

\section{Data}
\label{sec:data}
ALMA data from ATOMS survey (Project ID: 2019.1.00685.S; PI: Tie Liu) are used in this study. The region was observed on
3 November 2019 with ALMA 12-m array in band 3. In this study, we used 3 mm continuum and H$^{13}$CO$^{+}$ data with a beam size of $\sim2''\times 2''$ having rms noise of
0.2 mJy beam$^{-1}$ and 6.4 mJy beam$^{-1}$ per 0.122 MHz channel, respectively. More details of the ALMA observations and data reduction can be found in \citet{liu2020}.

We have also utilized the archival survey data from Two Micron All Sky Survey \citep[2MASS;][]{skrutskie2006}, {\it Spitzer Space Telescope} Galactic Legacy Infrared Mid-Plane Survey
Extraordinaire \citep[GLIMPSE;][]{benjamin2003}, Multiband Infrared Photometer for {\it Spitzer} \citep[MIPS][]{carey2005}, Wide-field Infrared Survey Explorer \citep[WISE;][]{wright2010}, and {\it Herschel}
Photoconductor Array Camera and Spectrometer \citep[PACS;][]{pilbratt2010}.

\section{Results}
\label{sec:results}
\subsection{Cores: Identification and Mass Estimation}
\label{core identification}
We identified the dust continuum cores in the ALMA 3 mm dust continuum image using the Python-based {\sc astrodendro} package \citep{rosolowsky2008} that uses the dendrogram algorithm. We considered
only those leaves (structures with no sub-structures) as cores that have area more than 33 pixels (i.e., size of the beam in pixel) in ALMA 3 mm dust continuum image and have signal-to-noise ratio of
more than 5. We identified four dust continuum cores (marked in Fig.\,\ref{RGB}(b)) and further refined their parameters using the CASA-{\sc imfit} task (listed in Table\,\ref{Tab-core}).

With the assumption that dust emission is optically thin, the masses of the four identified cores are estimated using the equation from \citet{hildebrand1983},
\begin{equation}
{M}_{\rm core} = \frac{\mathrm{F}_\nu\, D^2\, R_{gd}}{B_\nu (T_{\mathrm{dust}}) \,\kappa_\nu}
\end{equation}
where $F_\nu$ is integrated flux, D is distance to the region, R$_{\mathrm{gd}}$ is gas-to-dust ratio (adopted as 100), $T_{\mathrm{dust}}$ is dust temperature,
$ \kappa_\nu=10 (\nu /1.2\,\mathrm{THz})^\beta $ is the dust opacity where $\beta$ is the spectral index.
For dust temperature, we examined the {\it Herschel} dust temperature map \citep{marsh2017}. Fig.\,\ref{fig-ratio}(a) shows the 3-mm ALMA dust continuum map of the region overlaid
with {\it Herschel} dust temperature contours generated by \citet{marsh2017} using a Point Process Mapping (PPMAP) algorithm on multi-band {\it Herschel} dust continuum images. The final
resolution of the temperature map is $\sim$12$''$. As can be seen in Fig.\,\ref{fig-ratio}(a), all the identified dust continuum cores are located around T$_\mathrm{dust}$ contour of 21-22\,K.
Thus, we adapted an average $T_{\mathrm{dust}}$ of 21\,K for all the cores. The mass of the cores were estimated for two $\beta$ values of 1.5 and 2.0 and are listed in Table\,\ref{Tab-core}.

\begin{table}
\centering
\caption{Parameters of Identified Cores}
\bigskip

\addtolength{\tabcolsep}{-1.5pt}
\begin{tabular}{cccccc}
\hline
\textbf{Core\,[RA;Dec]}          & \textbf{Size}   & \textbf{Int. Flux} & \textbf{Peak Flux}       & \multicolumn{2}{c}{\textbf{Mass (M$_\odot$)}}                          \\
\textbf{(No\,[$^\circ$;$^\circ$])} &                 & \textbf{(mJy)}     & \textbf{(mJy beam$^{-1}$)} & \multicolumn{2}{c}{\textbf{for \textit{T}$_{\mathbf{dust}}$\,=\,21\,K}} \\
                                 &                 &                    &                           & \textbf{$\beta$=1.5} & \textbf{$\beta$=2.0} \\
\hline						
C1\,[253.1416;-44.6079] & 2.0$''\times$1.1$''$ & 2.7$\pm$0.1 & 1.4$\pm$0.1 &  9.9$\pm$0.2 &  34.3$\pm$0.8 \\ 
C2\,[253.1391;-44.6069] & 1.7$''\times$0.9$''$ & 2.1$\pm$0.1 & 1.3$\pm$0.1 &  7.8$\pm$0.3 &  27.1$\pm$1.0 \\
C3\,[253.1381;-44.6038] & 1.4$''\times$0.8$''$ &13.2$\pm$0.9 & 8.7$\pm$0.4 & 48.3$\pm$2.2 & 167.4$\pm$7.5 \\
C4\,[253.1387;-44.6029] & 2.6$''\times$1.7$''$ & 6.0$\pm$0.1 & 2.1$\pm$0.1 & 21.8$\pm$0.3 &  75.7$\pm$0.9 \\
\hline
\end{tabular}
\label{Tab-core}
\end{table}

\subsection{Young Stellar Objects}
We identifed Young Stellar Objects (YSOs) using three different color-color and color-magnitude criteria, and then merged them together. We first applied [3.6]-[24]/[3.6] color
criteria of  \citet{guieu2010} and \citet{rebull2011} to identify young sources. Then, we utilize MIR ([5.8]-[8.0])/([3.6]-[4.5]) color criteria given in \citet{gutermuth2009} to classify sources
that are not detected or saturated in 24 $\mu$m images. There are also sources that are not detected in 8$\mu$m band but detected in other IRAC bands, and for those we used ([3.6]-[4.5])/([4.5]-[5.8])
color-color scheme of \citet{hartmann2005,getman2007}. For unified YSOs catalog, we finally performed a cross-matching of all the YSOs candidates, and removed the duplicacy
\citep[see][for more details]{baug2016}. We found a total of 8 Class I and 5 Class II sources within the targeted regions. The positions of the identified YSOs are marked in Fig.\,\ref{fig-ratio}(b).

\subsection{Shocked Bubble}
\label{sec:Shocked bubble}
\begin{figure}[t]
\centering
\includegraphics[width=\textwidth]{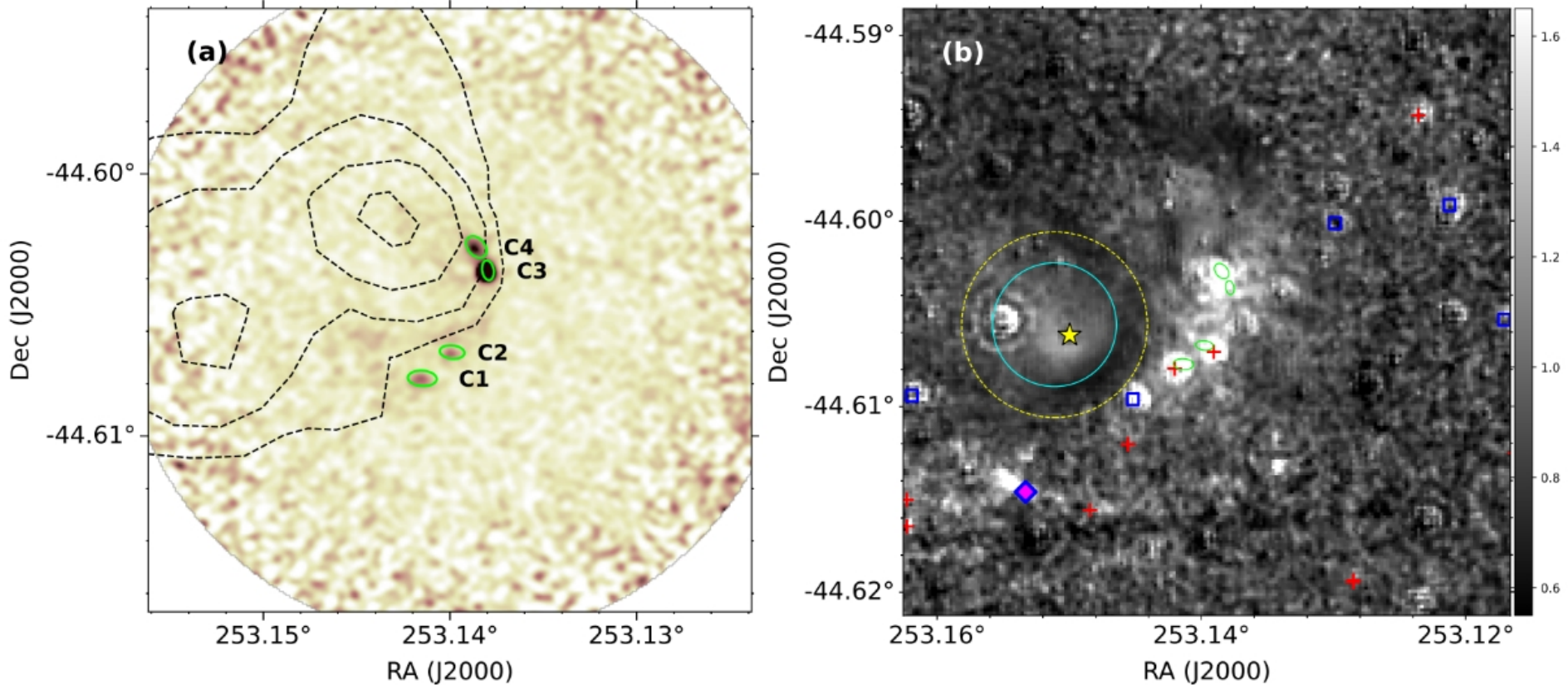}
\bigskip
\begin{minipage}{12cm}
\caption{(a) ALMA 3 mm continuum map of IRAS\,16489-4431 overlaid with {\it Herschel} dust temperature contours in black dashed lines drawn at 21, 22, 24 and 26\,K.
(b) Spitzer ratio map (Ch2/Ch1) of the target region. The position of Class I, Class II YSOs and the massive influencing star are marked by red pluses, blue squares 
and a yellow star, respectively. The yellow dashed circle shows a tentative outer (drawn by-eye) periphery of the shocked gas. The remaining symbols are the same
as they are in Fig.\,\ref{RGB}.  }
\label{fig-ratio}
\end{minipage}
\end{figure}

IRAS\,16489-4431 region contain a bubble at RA: 253.15107$^\circ$, Dec: -44.60557$^\circ$ having an angular radius of 0.2$'$ \citep{jayasinghe2019}. Ratio map of
{\it Spitzer}-IRAC images (e.g., Ch2/Ch1) can trace shocked regions as Ch2 includes shock excited H$_{2}$ emission and do not include any PAH features \citep{povich2007}.
Note that PAH emission can be triggered by both shock and ultra-violet radiation, but the  H$_{2}$ line in Ch2 is solely shock excited. Thus, a bright extended emission in the ratio
map can be inferred as a clear shocked region. On the otherhand, a dark extended emission in the ratio map is indicative of a tentative boundary of the shocked region.
A clear shocked morphology can be noted in the ratio map shown in Fig.\,\ref{fig-ratio}.

\begin{figure}[t]
\centering
\includegraphics[width=\textwidth]{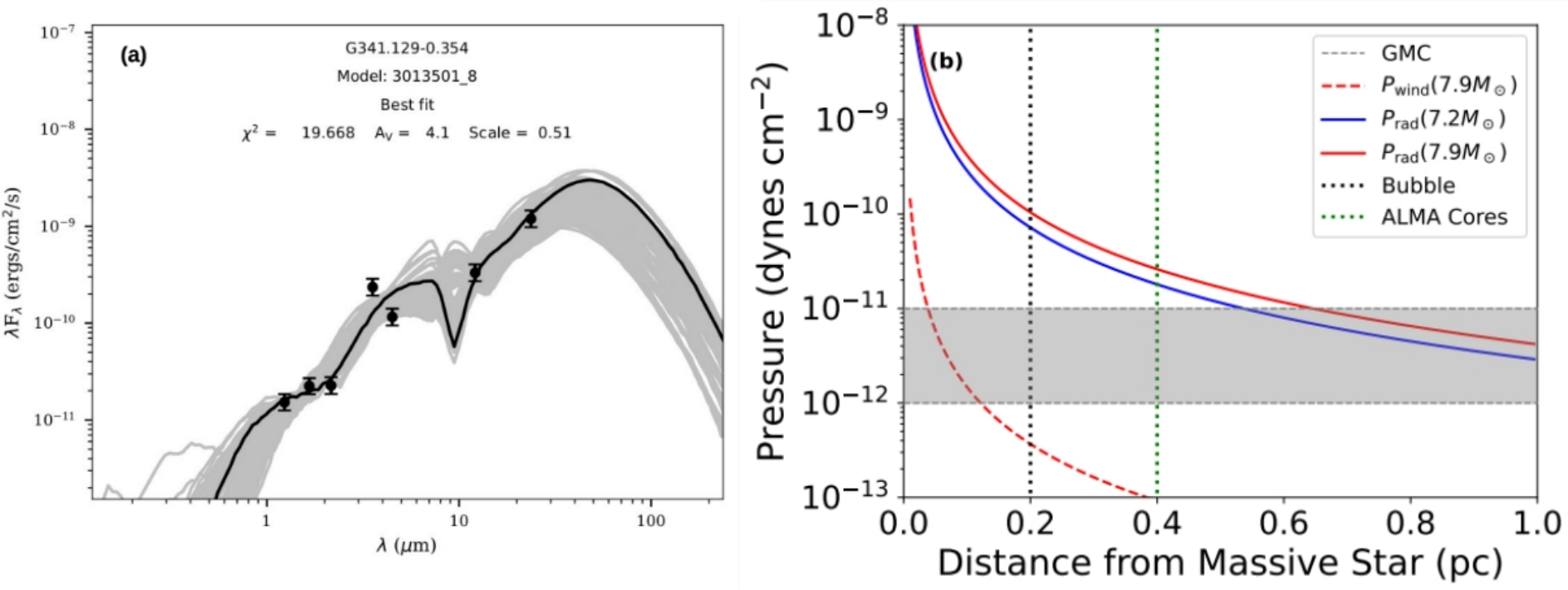}
\bigskip
\begin{minipage}{12cm}
\caption{(a) SED fit to the fluxes (black dots; with typical 20\% uncertainty) of the influencing source using the SED fitter tool of \citet{Robitaille2007}. 
The solid black curve and gray curves are the best fit model and other good fits, respectively. (b) Plot of different pressure components exerted by the influencing star as a function
of separation from the source. Radiation pressure for 7.9 M$_\odot$ and 7.2 M$_\odot$ stars are shown in red and blue solid curves. The wind pressure is shown in a red dashed curve.
The gray area marks the pressure exerted by a typical cold giant molecular cloud with temperature of 20 K and a particle density of 10$^{3}$--10$^{4}$ cm$^{-3}$.
The blue and green-dotted lines mark the average separation of the bubble edge and star-forming cores, respectively, from the influencing star.}
\label{fig-sed-pressure}
\end{minipage}
\end{figure}

We searched for a possible influencing source of the bubble. Photometric search revealed one potential source (RA: 253.14996$^\circ$, Dec: -44.60615$^\circ$) with
strong flux that could form the observed bubble. For confirmation, a fit to the observed fluxes (listed in Table\,\ref{Tab-source}) was performed using the SED-fitter tool of \citet{Robitaille2007}.
The input distance and interstellar visual extinction (A$_V$) ranges were set to 3.0-3.6 kpc and A$_V\sim$3-30, respectively. Among the fitted models
(see Fig.\,\ref{fig-sed-pressure}(a)), the models satisfying $\chi^2-\chi^2_\mathrm{best} \leq3$ were only considered in computing the mass of the source. 
The estimated mass of 7.9$\pm$0.7 M$_{\odot}$ is referring to a B2-B1 type star.

For a better knowledge of the feedback from the influencing star leading to the formation of the bubble, we computed different pressure components
following the similar method outlined in \citet{baug2019}. The pressure due to radiation was estimated using, P$_{rad}$=L$_{bol}$/4$\pi$cD$^2_s$, where L$_{bol}$ is the
bolometric luminosity of the star and D$_s$ is the separation from the source. The pressure due to stellar wind was computed using,
P$_\mathrm{w}$=$\dot{\mathrm{M}} V$/4$\pi$D$^2_s$, where $\dot{\mathrm{M}}$ is the mass-loss rate and $V$ is the velocity of stellar wind. We did not compute the pressure
for ionized gas as no radio continuum emission is detected within the bubble even in the recent CORNISH-South Survey \citep{irabor2023}.

The pressure components as a function of separation from the driving source is shown in Fig.\,\ref{fig-sed-pressure}(b). We showed the P$_{rad}$ for two masses, the lower
limit of the fitted mass (M$\sim$7.2 M$_\odot$) and 7.9 M$_\odot$. The corresponding bolometric luminosities of 2690 and 3900 L$_\odot$ are adopted from \citet{lang1999}.
For the calculation of P$_\mathrm{w}$, $V\sim$700 km s$^{-1}$ and $\dot{\mathrm{M}}$ of $\sim$10$^{-9.4}$ M$_\odot$ yr$^{-1}$ for a
typical B1V star were obtained from \citet{oskinova2011}. Fig.\,\ref{fig-sed-pressure}(b) shows that the radiation pressure dominates over stellar pressure and is
sufficiently higher than the typical pressure exerted by cool giant molecular cloud \citep[$\sim$10$^{-12}$--10$^{-11}$ dynes cm$^{-2}$ for a typical cloud temperature of
20 K and a particle density of 10$^{3}$--10$^{4}$ cm$^{-3}$;][]{dyson1980}.

\begin{table}
\caption{Photometric Magnitudes of the Influencing Source}
\bigskip

\addtolength{\tabcolsep}{-1.5pt}
\begin{tabular}{ccccccc}
\hline
\textbf{J$_{\mathbf{mag}}$} & \textbf{H$_{\mathbf{mag}}$} & \textbf{K$_{\mathbf{mag}}$} & \textbf{W1$_{\mathbf{mag}}$} & \textbf{W2$_{\mathbf{mag}}$} & \textbf{W3$_{\mathbf{mag}}$} & \textbf{W4$_{\mathbf{mag}}$}\\
\hline						
13.49$\pm$0.04 & 12.27$\pm$0.06 &11.51$\pm$0.05 &7.58$\pm$0.03 &7.46$\pm$0.02 &3.32$\pm$0.010 &-0.160$\pm$0.02\\ 
\hline
\end{tabular}
\label{Tab-source}
\end{table}

\section{Discussion}
\label{sec:discussion}
The primary aim of our study here is to assess the influence of a massive star on its surrounding gas. \citet{jayasinghe2019} had reported the presence of a shocked-bubble in this region.
Our analysis of IRAC-band images established the shocked nature of the bubble, and we also found that the bubble is developed by the strong radiation field
driven by a massive 7.9$\pm$0.7 M$_\odot$ star. In past decades, there have been several observational studies reported triggered star formation at the periphery of the MIR bubble
\citep{povich2008, smith2010, yuan2014, zhou2020}.

Based on the smoothed particle
hydrodynamic simulations of ionized-induced star formation in bound and unbound systems, \citet{dale2012, dale2013} argued that triggered star formation is most likely to occur at bubble
edges or pillars, whereas spontaneous star formation could occur in both bubble edge and within the bubble cavity. Also in \citet{dale2013} the authors suggested that the triggered stars
formed at the bubble wall are younger than that are formed spontaneously and have strong accretion. 

Our study shows the presence of collected materials (i.e., H$^{13}$CO$^{+}$ gas) towards the periphery of the bubble. Active star formation is also noted within
the compressed material around the bubble inferred from the presence of YSOs and star-forming cores. In addition, one of the star-forming cores is associated with bi-polar outflows, a
typical signature for a young protostellar core. Presence of Class {\sc ii} methanol maser source toward the southern periphery of the bubble is indicative of active massive star formation. 
In brief, all the signatures in the IRAS\,16489-4431 region hint toward an active star formation around the periphery of an expanding MIR bubble driven by a massive 7.9$\pm$0.7 M$_\odot$ star.

\section{Summary}
\label{sec:summary}
We performed a multiwavelength analysis of the local environment of the IRAS\,16489-4431 region associated with a MIR bubble. Analysis of {\it Spitzer}-IRAC band images revealed
shocked gas surrounding the edge of the bubble. We identified a possible massive 7.9$\pm$0.7 M$_\odot$ source which is driving the MIR bubble by its strong radiation pressure. ALMA
cold gas tracer showed evidence of collected gas around the bubble periphery where all the dense cores are located. One of these cores is associated with bi-polar
outflows indicative of active young star formation. The identified YSOs, specifically the Class I sources, are generally located around the edge of the shocked region. 
Overall, our analyses suggest that the radiation pressure from the 7.9$\pm$0.7 M$_\odot$ star has triggered the formation of the bubble and also might have influenced the star formation
activity around the bubble periphery.  

\begin{acknowledgments}
AH and TB thank the support by the S. N. Bose National Centre for Basic Sciences under the Department of Science and Technology, Govt. of India. AH also thanks
the CSIR-HRDG for the funding of the Junior Research Fellow.
This research makes use of astropy, a community-developed core python package for Astronomy \citep{astropy2018}. This paper makes use of the following ALMA data: 
ADS/JAO.ALMA 2019.1.00685.S and 2017.1.00545.S. ALMA is a partnership of ESO (representing its member states), NSF (USA), and NINS (Japan), together with NRC (Canada),
MOST and ASIAA (Taiwan), and KASI (Republic of Korea), in cooperation with the Republic of Chile. The Joint ALMA Observatory is operated by ESO, AUI/NRAO, and NAOJ.
\end{acknowledgments}

\begin{furtherinformation}

\begin{orcids}
\orcid{0009-0003-6633-525X}{Ariful}{Hoque}%
\orcid{0000-0003-0295-6586}{Tapas}{Baug}
\orcid{0000-0001-6725-0483}{Lokesh K.}{Dewangan}
\orcid{0000-0002-7237-3856}{Ke}{Wang}
\orcid{0000-0002-5286-2564}{Tie}{Liu}
\orcid{0000-0003-1457-0541}{Soumen}{Mondal}
\end{orcids}

\begin{authorcontributions}
All the authors of this paper have significantly contributed to the discussion and writing of this paper. 
\end{authorcontributions}

\begin{conflictsofinterest}
The authors declare no conflict of interest.
\end{conflictsofinterest}

\end{furtherinformation}

\bibliographystyle{bullsrsl-en}

\bibliography{S07-P06_HoqueA}

\begin{thebibliography}{39}
\providecommand{\natexlab}[1]{#1}
\providecommand{\url}[1]{#1}
\providecommand{\urlprefix}{URL }

\bibitem[{{Astropy Collaboration}(2018)}]{astropy2018}
{Astropy Collaboration} (2018) The astropy project: Building an open-science
  project and status of the v2.0 core package.
\newblock AJ, 156, 123.
\newblock \url{https://doi.org/10.3847/1538-3881/aabc4f}.

\bibitem[{Baug et~al.(2019)Baug, de~Grijs, Dewangan, Herczeg, Wang, Deng and
  Bhatt}]{baug2019}
Baug, T., de~Grijs, R., Dewangan, L.~K., Herczeg, D.~K., Gregory J.and~Ojha,
  Wang, K., Deng, L. and Bhatt, B.~C. (2019) Influence of wolf-rayet stars on
  surrounding star-forming molecular clouds.
\newblock ApJ, 885(1), 68.
\newblock \url{https://doi.org/10.3847/1538-4357/ab46be}.

\bibitem[{Baug et~al.(2016)Baug, Dewangan and Ninan}]{baug2016}
Baug, T., Dewangan, D.~K., L. K.and~Ojha and Ninan, J.~P. (2016) Star formation
  around mid-infrared bubble n37: Evidence of cloud-cloud collision.
\newblock ApJ, 833(1), 85.
\newblock \url{https://doi.org/10.3847/1538-4357/833/1/85}.

\bibitem[{Benjamin et~al.(2003)Benjamin, Churchwell, Babler, Bania, Clemens,
  Cohen, Dickey, Indebetouw, Jackson, Kobulnicky, Lazarian, Marston, Mathis,
  Meade, Seager, Stolovy, Watson, Whitney, Wolff and Wolfire}]{benjamin2003}
Benjamin, R.~A., Churchwell, E., Babler, B.~L., Bania, T.~M., Clemens, D.~P.,
  Cohen, M., Dickey, J.~M., Indebetouw, R., Jackson, J.~M., Kobulnicky, H.~A.,
  Lazarian, A., Marston, A.~P., Mathis, J.~S., Meade, M.~R., Seager, S.,
  Stolovy, S.~R., Watson, C., Whitney, B.~A., Wolff, M.~J. and Wolfire, M.~G.
  (2003) Glimpse. i. an sirtf legacy project to map the inner galaxy.
\newblock MNRAS, 115, 953--964.
\newblock \url{https://doi.org/10.1086/376696}.

\bibitem[{Carey et~al.(2005)Carey, Noriega-Crespo, Price, Padgett, Kraemer,
  Indebetouw, Mizuno, Ali, Berriman, Boulanger, Cutri, Ingalls, Kuchar, Latter,
  Marleau, Miville-Deschenes, Molinari, Rebull and Testi}]{carey2005}
Carey, S.~J., Noriega-Crespo, A., Price, S.~D., Padgett, D.~L., Kraemer, K.~E.,
  Indebetouw, R., Mizuno, D.~R., Ali, B., Berriman, G.~B., Boulanger, F.,
  Cutri, R.~M., Ingalls, J.~G., Kuchar, T.~A., Latter, W.~B., Marleau, F.~R.,
  Miville-Deschenes, M.~A., Molinari, S., Rebull, L.~M. and Testi, L. (2005)
  Mipsgal: A survey of the inner galactic plane at 24 and 70 microns, survey
  strategy and early results.
\newblock BAAS, 37, 63.33.

\bibitem[{Churchwell et~al.(2006)Churchwell, Povich, Allen, Taylor, Meade,
  Babler, Indebetouw, Watson, Whitney, Wolfire, Bania, Benjamin, Clemens,
  Cohen, Cyganowski, Jackson, Kobulnicky, Mathis, Mercer, Stolovy, Uzpen,
  Watson and Wolff}]{churchwell2006}
Churchwell, E., Povich, M.~S., Allen, D., Taylor, M.~G., Meade, M.~R., Babler,
  B.~L., Indebetouw, R., Watson, C., Whitney, B.~A., Wolfire, M.~G., Bania,
  T.~M., Benjamin, R.~A., Clemens, D.~P., Cohen, M., Cyganowski, C.~J.,
  Jackson, J.~M., Kobulnicky, H.~A., Mathis, J.~S., Mercer, E.~P., Stolovy,
  S.~R., Uzpen, B., Watson, D.~F. and Wolff, M.~J. (2006) The bubbling galactic
  disk.
\newblock ApJ, 649(2), 759--778.
\newblock \url{https://doi.org/10.1086/507015}.

\bibitem[{Churchwell et~al.(2007)Churchwell, Watson, Povich, Taylor, Babler,
  Meade, Benjamin, Indebetouw and Whitney}]{churchwell2007}
Churchwell, E., Watson, D.~F., Povich, M.~S., Taylor, M.~G., Babler, B.~L.,
  Meade, M.~R., Benjamin, R.~A., Indebetouw, R. and Whitney, B.~A. (2007) The
  bubbling galactic disk. ii. the inner 20$^\circ$.
\newblock ApJ, 670(1), 428--441.
\newblock \url{https://doi.org/10.1086/521646}.

\bibitem[{{Dale} et~al.(2012){Dale}, {Ercolano} and {Bonnell}}]{dale2012}
{Dale}, J.~E., {Ercolano}, B. and {Bonnell}, I.~A. (2012) {Ionization-induced
  star formation - IV. Triggering in bound clusters}.
\newblock MNRAS, 427(4), 2852--2865.
\newblock \url{https://doi.org/10.1111/j.1365-2966.2012.22104.x}.

\bibitem[{{Dale} et~al.(2013){Dale}, {Ercolano} and {Bonnell}}]{dale2013}
{Dale}, J.~E., {Ercolano}, B. and {Bonnell}, I.~A. (2013) {Ionization-induced
  star formation - V. Triggering in partially unbound clusters}.
\newblock MNRAS, 431(2), 1062--1076.
\newblock \url{https://doi.org/10.1093/mnras/stt236}.

\bibitem[{Das et~al.(2017)Das, Tej, Vig, Liu, Liu, Ishwara~Chandra and
  Ghosh}]{das2017}
Das, S.~R., Tej, A., Vig, S., Liu, H.-L., Liu, T., Ishwara~Chandra, C.~H. and
  Ghosh, S.~K. (2017) Infrared dust bubble cs51 and its interaction with the
  surrounding interstellar medium.
\newblock MNRAS, 472(4), 4750--4768.
\newblock \url{https://doi.org/10.1093/mnras/stx2290}.

\bibitem[{Deharveng et~al.(2010)Deharveng, Schuller, Anderson, Zavagno,
  Wyrowski, Menten, Bronfman, Testi, Walmsley and Wienen}]{deharveng2010}
Deharveng, L., Schuller, F., Anderson, L.~D., Zavagno, A., Wyrowski, F.,
  Menten, K.~M., Bronfman, L., Testi, L., Walmsley, C.~M. and Wienen, M. (2010)
  A gallery of bubbles. the nature of the bubbles observed by spitzer and what
  atlasgal tells us about the surrounding neutral material.
\newblock A\&A, 523(A6).
\newblock \url{https://doi.org/10.1051/0004-6361/201014422}.

\bibitem[{Dewangan et~al.(2020)Dewangan, Baug, Pirogov and Ojha}]{dewangan2020}
Dewangan, L.~K., Baug, T., Pirogov, L.~E. and Ojha, D.~K. (2020) Investigating
  the physical conditions in extended system hosting mid-infrared bubble n14.
\newblock ApJ, 898(1), 41.
\newblock \url{https://doi.org/10.3847/1538-4357/ab964c}.

\bibitem[{{Dyson} and {Williams}(1980)}]{dyson1980}
{Dyson}, J.~E. and {Williams}, D.~A. (1980) {Physics of the interstellar
  medium}.
\newblock Manchester: University Press.

\bibitem[{Elmegreen and Lada(1977)}]{elmegreen1977}
Elmegreen, B.~G. and Lada, C.~J. (1977) Sequential formation of subgroups in ob
  associations.
\newblock ApJ, 214, 725--741.
\newblock \url{https://doi.org/10.1086/155302}.

\bibitem[{Getman et~al.(2007)Getman, Feigelson, Garmire, Broos and
  Wang}]{getman2007}
Getman, K.~V., Feigelson, E.~D., Garmire, G., Broos, P. and Wang, J. (2007)
  X-ray study of triggered star formation and protostars in ic 1396n.
\newblock ApJ, 654(1), 316--337.
\newblock \url{https://doi.org/10.1086/509112}.

\bibitem[{Guieu et~al.(2010)Guieu, Rebull, Stauffer, Vrba, Noriega-Crespo,
  Spuck, Roelofsen~Moody, Sepul, Cole, Flagey and Laher}]{guieu2010}
Guieu, S., Rebull, L.~M., Stauffer, J.~R., Vrba, F.~J., Noriega-Crespo, A.,
  Spuck, T., Roelofsen~Moody, T., Sepul, Cole, D.~M., Flagey, N. and Laher, R.
  (2010) Spitzer observations of ic 2118.
\newblock ApJ, 720(1), 46--63.
\newblock \url{https://doi.org/10.1088/0004-637X/720/1/46}.

\bibitem[{Gutermuth et~al.(2009)Gutermuth, Megeath, Myers, Allen, Pipher and
  Fazio}]{gutermuth2009}
Gutermuth, R.~A., Megeath, S.~T., Myers, P.~C., Allen, L.~E., Pipher, J.~L. and
  Fazio, G.~G. (2009) A spitzer survey of young stellar clusters within one
  kiloparsec of the sun: Cluster core extraction and basic structural analysis.
\newblock ApJS, 184(1), 18--83.
\newblock \url{https://doi.org/10.1088/0067-0049/184/1/18}.

\bibitem[{Hartmann et~al.(2005)Hartmann, Megeath, Allen, Luhman, Calvet,
  D'Alessio, Franco-Hernandez and Fazio}]{hartmann2005}
Hartmann, L., Megeath, S.~T., Allen, L., Luhman, K., Calvet, N., D'Alessio, P.,
  Franco-Hernandez, R. and Fazio, G. (2005) Irac observations of taurus
  pre-main-sequence stars.
\newblock ApJ, 629(2), 881--896.
\newblock \url{https://doi.org/10.1086/431472}.

\bibitem[{Hildebrand(1983)}]{hildebrand1983}
Hildebrand, R.~H. (1983) The determination of cloud masses and dust
  characteristics from submillimetre thermal emission.
\newblock QJRAS, 24(3), 267--282.

\bibitem[{{Irabor} et~al.(2023){Irabor}, {Hoare}, {Burton}, {Cotton},
  {Diamond}, {Dougherty}, {Ellingsen}, {Fender}, {Fuller}, {Garrington},
  {Goldsmith}, {Green}, {Gunn}, {Jackson}, {Kurtz}, {Lumsden}, {Marti},
  {McDonald}, {Molinari}, {Moore}, {Mutale}, {Muxlow}, {O'Brien}, {Oudmaijer},
  {Paladini}, {Pandian}, {Paredes}, {Richards}, {Sanchez-Monge}, {Spencer},
  {Thompson}, {Umana}, {Urquhart}, {Wieringa} and {Zijlstra}}]{irabor2023}
{Irabor}, T., {Hoare}, M.~G., {Burton}, M., {Cotton}, W.~D., {Diamond}, P.,
  {Dougherty}, S., {Ellingsen}, S.~P., {Fender}, R., {Fuller}, G.~A.,
  {Garrington}, S., {Goldsmith}, P.~F., {Green}, J., {Gunn}, A.~G., {Jackson},
  J., {Kurtz}, S., {Lumsden}, S.~L., {Marti}, J., {McDonald}, I., {Molinari},
  S., {Moore}, T.~J., {Mutale}, M., {Muxlow}, T., {O'Brien}, T., {Oudmaijer},
  R.~D., {Paladini}, R., {Pandian}, J.~D., {Paredes}, J.~M., {Richards},
  A.~M.~S., {Sanchez-Monge}, A., {Spencer}, R., {Thompson}, M.~A., {Umana}, G.,
  {Urquhart}, J.~S., {Wieringa}, M. and {Zijlstra}, A. (2023) {The coordinated
  radio and infrared survey for high-mass star formation - V. The CORNISH-South
  survey and catalogue}.
\newblock MNRAS, 520(1), 1073--1091.
\newblock \url{https://doi.org/10.1093/mnras/stad005}.

\bibitem[{Jayasinghe et~al.(2019)Jayasinghe, Dixon, Povich, Binder, Velasco,
  Lepore, Xu, Offner, Kobulnicky, Anderson, Kendrew and
  Simpson}]{jayasinghe2019}
Jayasinghe, T., Dixon, D., Povich, M.~S., Binder, B., Velasco, J., Lepore,
  D.~M., Xu, D., Offner, S., Kobulnicky, H.~A., Anderson, L.~D., Kendrew, S.
  and Simpson, R.~J. (2019) The milky way project second data release: bubbles
  and bow shocks.
\newblock MNRAS, 488(1), 1141--1165.
\newblock \url{https://doi.org/10.1093/mnras/stz1738}.

\bibitem[{{Lang}(1999)}]{lang1999}
{Lang}, K.~R. (1999) {Astrophysical formulae}.
\newblock New York : Springer.

\bibitem[{Liu et~al.(2020)Liu, Evans, Kim, Goldsmith, Liu, Zhang, Tatematsu,
  Wang, Juvela, Bronfman, Cunningham, Garay, Hirota, Lee, Kang, Li, Li,
  Mardones, Qin, Ristorcelli, Tej, Toth, Wu, Wu, Yi, Yun, Liu, Peng, Li, Li,
  Lee, Shen, Baug, Wang, Zhang, Issac, Zhu, Luo, Soam, Liu, Xu, Wang, Zhang,
  Ren and Zhang}]{liu2020}
Liu, T., Evans, N.~J., Kim, K., Goldsmith, P.~F., Liu, S., Zhang, Q.,
  Tatematsu, K., Wang, K., Juvela, M., Bronfman, L., Cunningham, M.~R., Garay,
  G., Hirota, T., Lee, J., Kang, S., Li, D., Li, P., Mardones, D., Qin, S.,
  Ristorcelli, I., Tej, A., Toth, L.~V., Wu, J., Wu, Y., Yi, H., Yun, H., Liu,
  H., Peng, Y., Li, J., Li, S., Lee, C.~W., Shen, Z., Baug, T., Wang, J.,
  Zhang, Y., Issac, N., Zhu, F., Luo, Q., Soam, A., Liu, X., Xu, F., Wang, Y.,
  Zhang, C., Ren, Z. and Zhang, C. (2020) Atoms: Alma three-millimeter
  observations of massive star-forming regions - i. survey description and a
  first look at g9.62+0.19.
\newblock MNRAS, 496(3), 2790--2820.
\newblock \url{https://doi.org/10.1093/mnras/staa1577}.

\bibitem[{{Marsh} et~al.(2017){Marsh}, {Whitworth}, {Lomax}, {Ragan},
  {Becciani}, {Cambr{\'e}sy}, {Di Giorgio}, {Eden}, {Elia}, {Kacsuk},
  {Molinari}, {Palmeirim}, {Pezzuto}, {Schneider}, {Sciacca} and
  {Vitello}}]{marsh2017}
{Marsh}, K.~A., {Whitworth}, A.~P., {Lomax}, O., {Ragan}, S.~E., {Becciani},
  U., {Cambr{\'e}sy}, L., {Di Giorgio}, A., {Eden}, D., {Elia}, D., {Kacsuk},
  P., {Molinari}, S., {Palmeirim}, P., {Pezzuto}, S., {Schneider}, N.,
  {Sciacca}, E. and {Vitello}, F. (2017) {Multitemperature mapping of dust
  structures throughout the Galactic Plane using the PPMAP tool with Herschel
  Hi-GAL data}.
\newblock MNRAS, 471(3), 2730--2742.
\newblock \url{https://doi.org/10.1093/mnras/stx1723}.

\bibitem[{Oskinova et~al.(2011)Oskinova, Todt, Ignace, Brown, Cassinelli and
  Hamann}]{oskinova2011}
Oskinova, L.~M., Todt, H., Ignace, R., Brown, J.~C., Cassinelli, J.~P. and
  Hamann, W.~R. (2011) Early magnetic b-type stars: X-ray emission and wind
  properties.
\newblock MNRAS, 416(2), 1456--1474.
\newblock \url{https://doi.org/10.1111/j.1365-2966.2011.19143.x}.

\bibitem[{Pilbratt et~al.(2010)Pilbratt, Riedinger, Passvogel, Crone, Doyle,
  Gageur, Heras, Jewell, Metcalfe, Ott and Schmidt}]{pilbratt2010}
Pilbratt, G.~L., Riedinger, J.~R., Passvogel, T., Crone, G., Doyle, D., Gageur,
  U., Heras, A.~M., Jewell, C., Metcalfe, L., Ott, S. and Schmidt, M. (2010)
  Herschel space observatory. an esa facility for far-infrared and
  submillimetre astronomy.
\newblock A\&A, 518, L1.
\newblock \url{https://doi.org/10.1051/0004-6361/201014759}.

\bibitem[{Povich et~al.(2008)Povich, Benjamin, Whitney, Babler, Indebetouw,
  Meade and Churchwell}]{povich2008}
Povich, M.~S., Benjamin, R.~A., Whitney, B.~A., Babler, B.~L., Indebetouw, R.,
  Meade, M.~R. and Churchwell, E. (2008) Interstellar weather vanes: Glimpse
  mid-infrared stellar wind bow shocks in m17 and rcw 49.
\newblock ApJ, 689(1), 242--248.
\newblock \url{https://doi.org/10.1086/431472}.

\bibitem[{Povich et~al.(2007)Povich, Stone, Churchwell, Zweibel, Wolfire,
  Babler, Indebetouw, Meade and Whitney}]{povich2007}
Povich, M.~S., Stone, J.~M., Churchwell, E., Zweibel, E.~G., Wolfire, M.~G.,
  Babler, B.~L., Indebetouw, R., Meade, M.~R. and Whitney, B.~A. (2007) A
  multiwavelength study of m17: The spectral energy distribution and pah
  emission morphology of a massive star formation region.
\newblock ApJ, 660(1), 346--362.
\newblock \url{https://doi.org/10.1086/513073}.

\bibitem[{Rebull et~al.(2011)Rebull, Johnson, Hoette, Kim, Laine, Foster,
  Laher, Legassie, Mallory and McCarron}]{rebull2011}
Rebull, L.~M., Johnson, C.~H., Hoette, V., Kim, J.~S., Laine, S., Foster, M.,
  Laher, R., Legassie, M., Mallory, C.~R. and McCarron, K. (2011) New young
  star candidates in cg4 and sa101.
\newblock AJ, 142(1), 25--46.
\newblock \url{https://doi.org/10.1088/0004-6256/142/1/25}.

\bibitem[{Robitaille et~al.(2007)Robitaille, Whitney, Indebetouw and
  Wood}]{Robitaille2007}
Robitaille, T.~P., Whitney, B.~A., Indebetouw, R. and Wood, K. (2007)
  Interpreting spectral energy distributions from young stellar objects. ii.
  fitting observed seds using a large grid of precomputed models.
\newblock ApJS, 169(2), 328--352.
\newblock \url{https://doi.org/10.1086/512039}.

\bibitem[{Rosolowsky et~al.(2008)Rosolowsky, Pineda, Kauffmann and
  Goodman}]{rosolowsky2008}
Rosolowsky, E.~W., Pineda, J.~E., Kauffmann, J. and Goodman, A.~A. (2008)
  Structural analysis of molecular clouds: Dendrograms.
\newblock ApJ, 679(2), 1338--1351.
\newblock \url{https://doi.org/10.1086/587685}.

\bibitem[{Skrutskie et~al.(2006)Skrutskie, Cutri, Stiening, Weinberg,
  Schneider, Carpenter, Beichman, Capps, Chester, Elias, Huchra, Liebert,
  Lonsdale, Monet, Price, Seitzer, Jarrett, Kirkpatrick, Gizis, Howard, Evans,
  Fowler, Fullmer, Hurt, Light, Kopan, Marsh, McCallon, Tam, Van~Dyk and
  Wheelock}]{skrutskie2006}
Skrutskie, M.~F., Cutri, R.~M., Stiening, R., Weinberg, M.~D., Schneider, S.,
  Carpenter, J.~M., Beichman, C., Capps, R., Chester, T., Elias, J., Huchra,
  J., Liebert, J., Lonsdale, C., Monet, D.~G., Price, S., Seitzer, P., Jarrett,
  T., Kirkpatrick, J.~D., Gizis, J.~E., Howard, E., Evans, T., Fowler, J.,
  Fullmer, L., Hurt, R., Light, R., Kopan, E.~L., Marsh, K.~A., McCallon,
  H.~L., Tam, R., Van~Dyk, S. and Wheelock, S. (2006) The two micron all sky
  survey (2mass).
\newblock AJ, 131(2), 1163--1183.
\newblock \url{https://doi.org/10.1086/498708}.

\bibitem[{Smith et~al.(2010)Smith, Povich, Whitney, Churchwell, Babler, Meade,
  Bally, Gehrz, Robitaille and Stassun}]{smith2010}
Smith, N., Povich, M.~S., Whitney, B.~A., Churchwell, E., Babler, B.~L., Meade,
  M.~R., Bally, J., Gehrz, R.~D., Robitaille, T.~P. and Stassun, K.~G. (2010)
  Spitzer space telescope observations of the carina nebula: the steady march
  of feedback-driven star formation.
\newblock MNRAS, 402(2), 952--974.
\newblock \url{https://doi.org/10.1111/j.1365-2966.2010.16792.x}.

\bibitem[{Stromgren(1939)}]{stromgren1939}
Stromgren, B. (1939) The physical state of interstellar hydrogen.
\newblock ApJ, 89, 526--547.
\newblock \url{https://doi.org/10.1086/144074}.

\bibitem[{Urquhart et~al.(2018)Urquhart, Konig, Giannetti, Leurini, Moore,
  Eden, Pillai, Thompson, Braiding, Burton, Csengeri, Dempsey, Figura,
  Froebrich, Menten, Schuller and Smith}]{urquhart2018}
Urquhart, J.~S., Konig, C., Giannetti, A., Leurini, S., Moore, T. J.~T., Eden,
  D.~J., Pillai, T., Thompson, M.~A., Braiding, C., Burton, M.~G., Csengeri,
  T., Dempsey, J.~T., Figura, C., Froebrich, D., Menten, K.~M., Schuller, F.
  and Smith, M.~D., M. D.and~Wyrowski (2018) Atlasgal-properties of a complete
  sample of galactic clumps.
\newblock MNRAS, 473(1), 1059--1102.
\newblock \url{https://doi.org/10.1093/mnras/stx2258}.

\bibitem[{Wright et~al.(2010)Wright, Eisenhardt, Mainzer, Ressler, Cutri,
  Jarrett, Kirkpatrick, Padgett, McMillan, Skrutskie, Stanford, Cohen, Walker,
  Mather, Leisawitz, Gautier, McLean, Benford, Lonsdale, Blain, Mendez, Irace,
  Duval, Liu, Royer, Heinrichsen, Howard, Shannon, Kendall, Walsh, Larsen,
  Cardon, Schick, Schwalm, Abid, Fabinsky, Naes and Tsai}]{wright2010}
Wright, E.~L., Eisenhardt, P. R.~M., Mainzer, A.~K., Ressler, M.~E., Cutri,
  R.~M., Jarrett, T., Kirkpatrick, J.~D., Padgett, D., McMillan, R.~S.,
  Skrutskie, M., Stanford, S.~A., Cohen, M., Walker, R.~G., Mather, J.~C.,
  Leisawitz, D., Gautier, T.~N., McLean, I., Benford, D., Lonsdale, C.~J.,
  Blain, A., Mendez, B., Irace, W.~R., Duval, V., Liu, F., Royer, D.,
  Heinrichsen, I., Howard, J., Shannon, M., Kendall, M., Walsh, A.~L., Larsen,
  M., Cardon, J.~G., Schick, S., Schwalm, M., Abid, M., Fabinsky, B., Naes, L.
  and Tsai, C. (2010) The wide-field infrared survey explorer (wise): Mission
  description and initial on-orbit performance.
\newblock AJ, 140(6), 1868--1881.
\newblock \url{https://doi.org/10.1088/0004-6256/140/6/1868}.

\bibitem[{{Yuan} et~al.(2014){Yuan}, {Wu}, {Li} and {Liu}}]{yuan2014}
{Yuan}, J.-H., {Wu}, Y., {Li}, J.~Z. and {Liu}, H. (2014) {Expanding Shell and
  Star Formation in the Infrared Dust Bubble N6}.
\newblock ApJ, 797(1), 40.
\newblock \url{https://doi.org/10.1088/0004-637X/797/1/40}.

\bibitem[{{Zhou} et~al.(2020){Zhou}, {Zhou}, {Esimbek}, {Baan}, {Wu}, {Ji},
  {He}, {Li}, {Sailanbek}, {Komesh} and {Tang}}]{zhou2020}
{Zhou}, J., {Zhou}, D., {Esimbek}, J., {Baan}, W., {Wu}, G., {Ji}, W., {He},
  Y., {Li}, D., {Sailanbek}, S., {Komesh}, T. and {Tang}, X. (2020)
  {G15.684-0.29: One of the Largest Galactic Infrared Bubbles Showing Strong
  Evidence of Triggered Star Formation}.
\newblock ApJ, 897(1), 74.
\newblock \url{https://doi.org/10.3847/1538-4357/ab94c0}.

\bibitem[{Zinnecker and Yorke(2007)}]{zinnecker2007}
Zinnecker, H. and Yorke, H.~W. (2007) Toward understanding massive star
  formation.
\newblock ARA\&A, 45(1), 481--563.
\newblock \url{https://doi.org/10.1146/annurev.astro.44.051905.092549}.

\end{thebibliography}

\end{document}